# Electrical and Optical Properties of Heavily Ge-Doped AlGaN

R. Blasco[1,2], A. Ajay[1], E. Robin[1], C Bougerol[3], K. Lorentz[4], L. C. Alves[5], I. Mouton[1], L Amichi[1], A. Grenier[6], and E. Monroy[1]

[1] Univ. Grenoble-Alpes, CEA, INAC, 17 av. des Martyrs, 38000 Grenoble, France.

[2] Grupo de Ingeniería Fotónica, Dept. de Electrónica, Universidad de Alcalá, 28871 Alcalá de Henares, Spain.

[3] Univ. Grenoble-Alpes, Institut Néel-CNRS, 25 av. des Martyrs, 38000 Grenoble, France.

[4] INESC-MN, IPFN, Instituto Superior Técnico (IST), Campus Tecnológico e Nuclear, 2695-066 Bobadela LRS, Portugal.

[5] C2TN, Instituto Superior Técnico (IST), Campus Tecnológico e Nuclear, 2695-066 Bobadela LRS, Portugal.

[6] CEA-LETI, 17 av. des Martyrs, 38054 Grenoble, France.

## ABSTRACT

We report the effect of germanium as n-type dopant on the electrical and optical properties of $Al_xGa_{1-x}N$ layers grown by plasma-assisted molecular-beam epitaxy. The Al content has been varied from x = 0 to 0.66, confirmed by Rutherford backscattering spectrometry, and the Ge concentration was increased up to [Ge] = $1\times10^{21}$ cm$^{-3}$. Even at these high doping levels (> 1% atomic fraction) Ge does not induce any structural degradation in $Al_xGa_{1-x}N$ layers with x < 0.15. However, for higher Al compositions, clustering of Ge forming crystallites were observed. Hall effect measurements show a gradual decrease of the carrier concentration when increasing the Al mole fraction, which is already noticeable in samples with x = 0.24. Samples with x = 0.64-0.66 remain conductive (σ = 0.8-0.3 $\Omega^{-1}$cm$^{-1}$), but the donor activation rate drops to around 0.1% (carrier concentration around $1\times10^{18}$ cm$^{-3}$ for [Ge] ≈ $1\times10^{21}$ cm$^{-3}$). From the optical point of view, the low temperature photoluminescence is dominated by the band-to-band emission, which show only spectral shift and broadening associated to the Burstein-Moss effect. The evolution of the photoluminescence peak position with temperature shows that the free carriers due to Ge doping can efficiently screen the potential fluctuations induced by alloy disorder.

## 1. Introduction

Although silicon has long been the preferred n-type dopant for AlGaN, germanium is currently under consideration, particularly in those applications that require dopant concentrations around or higher than $10^{19}$ $cm^{-3}$ [1–4]. In the case of AlGaN planar layers, high Si doping levels contribute to edge-type dislocation climb and induce tensile stress [4–7], which leads to a degradation of the surface morphology and favors crack propagation. In the case of nanowires, Si tends to migrate towards the sidewalls [8], and high silicon concentration degrades the nanowire morphology [9].

On the other hand, achieving highly conductive Si-doped $Al_xGa_{1-x}N$ for $x > 0.70$ has proven difficult due to a sharp increase in the donor activation energy [10–12] and resistivity [13]. Carrier compensation by deep level defects, including deep Si DX centers, has often been speculated. DX centers are formed when a shallow donor impurity undergoes a large bond-rupturing displacement and becomes a deep acceptor by trapping electrons, which leads to a dramatic drop of the free carrier density. Different calculations support that Si forms a deep DX center in $Al_xGa_{1-x}N$ [14–17]. Park and Chadi [16] expected the onset of DX behavior for Si to occur at $x > 0.24$, whereas Boguslawski and Bernholc [13] predicted that Si-related DX centers are stable at $x > 0.60$, and Gordon et al. [15] obtained an onset of DX transition for $x = 0.94$. Experimental results are also contradictory, e.g. some studies suggested Si to be a DX center in $Al_xGa_{1-x}N$ for $x > 0.5$ [18], others for $x \geq 0.84$ [19], and some show indications of self-compensation for relatively high doping levels ([Si] > $3\times10^{19}$ $cm^{-3}$) for $x \geq 0.42$ [20]. The dispersion of the experimental results suggests that the electrical properties are not only related to the behavior of Si donors, but also strongly influenced by the growth-dependent structural characteristics of the samples.

In this framework, an alternative approach to obtain high carrier concentrations in AlGaN is to exploit the effect of spontaneous polarization gradients in graded-alloy AlGaN layers [21,22]. However, this polarization-induced doping is bound to a gradient of lattice parameter and band gap, which imposes serious limitations to the device design.

Germanium is currently being reconsidered as a n-type dopant for AlGaN, particularly for doping concentrations above the Mott transition ($\approx 10^{19}$ $cm^{-3}$) [4,6]. Ge is a shallow donor in GaN with an activation energy of 31.1 meV [23]. Looking at the Ge-N and Si-N bond lengths (around 1.71-1.77 Å and 1.89-2.03 Å, respectively) in comparison with the Ga-N and Al-N bonds (around 1.95 Å and 1.89 Å, respectively), Ge should occupy the Ga or Al lattice sites in AlGaN causing far less lattice distortion than Si. As experimental evidences, the lower tensile strain in Ge-doped GaN in comparison with Si-doped GaN was demonstrated by Fritze et al. [6], and growth of Ge-doped GaN nanowires with metallic behavior without deformation of the nanowire morphology has been demonstrated [24]. However, there is very little experimental information about the behavior of Ge in AlGaN. From the theoretical viewpoint, Gordon et al. [15] predicted the onset of the DX transition for Ge in $Al_xGa_{1-x}N$ at $x = 0.52$, but there is no experimental evidence at this point. Therefore, there is an interest to perform experimental studies of Ge as a dopant in AlGaN.

In this work, we report the effect of germanium as n-type dopant in $Al_xGa_{1-x}N$ layers grown by plasma assisted molecular-beam epitaxy. Various series of Ge-doped $Al_xGa_{1-x}N$ samples were grown, first varying the Ge concentration in the range of $10^{19}$ to $10^{21}$ cm$^{-3}$ for relatively low Al compositions (x = 0.12, 0.24 and 0.36), and then with fixed Ge density ([Ge] = $1\times10^{21}$ cm$^{-3}$) increasing the Al composition up to 0.66. In this latter series with [Ge] = $1\times10^{21}$ cm$^{-3}$, Hall Effect measurements at room temperature show a gradual decrease of the carrier concentration when increasing the Al mole fraction. Structural characterization reveals clustering of Ge for alloys with x > 0.15, which might explain the reduction of the donor activation rate.

## 2. Methods

Ge-doped $Al_xGa_{1-x}N$ thin films with thickness of 675 nm were grown by plasma-assisted molecular beam epitaxy (PAMBE) on 1-μm-thick AlN-on-sapphire templates. The growth was performed under slightly metal-rich conditions, with $(\Phi_{Al} + \Phi_{Ga})/\Phi_N \approx 1.1$. The substrate temperature was kept in the 710-720°C range. The aluminum cell temperature was fixed so that the Al flux is $\Phi_{Al} = x\Phi_N$ where x was the desired Al mole fraction and $\Phi_N \approx 0.5$ monolayers per second (ML/s) was the active nitrogen flux. The metal excess is provided by the Ga flux. The Ga cell temperature was tuned to obtain stable growth conditions, with a self-regulated Ga layer covering the growth front [25,26]. This situation can be monitored in real time by verifying that the reflection high-energy electron diffraction (RHEED) intensity remains constant during the growth process, while maintaining a streaky RHEED pattern. The Al mole fraction was increased progressively up to 0.66 and the Ge cell temperatures were $T_{Ge}$ = 840°C, 928°C, or 1011°C, which lead to Ge concentrations around $10^{19}$, $1.5\times10^{20}$ and $1.3\times10^{21}$ cm$^{-3}$, respectively. The list of samples under study is summarized in Table I.

Rutherford backscattering spectrometry (RBS) measurements were performed using a collimated 1.8 MeV $^4He^+$ ion beam and a silicon surface barrier detector at a scattering angle of 160°. To extract chemical composition and thickness, spectra were fitted using the NDF code [27].

The structural quality of the layers was studied by x-ray diffraction (XRD) using a Rigaku SmartLab x-ray diffractometer using a 4-bounce Ge(220) monochromator and a Si compensator.

For particle induced x-ray emission (PIXE) analysis, a 2 MeV focused proton beam ($3\times4\ \mu m^2$) was raster scanned over the samples, and the produced x-ray signal was collected with a 145 eV resolution Si(Li) detector positioned at 45° to the beam direction. The OMDAQ 2007 software was used to deconvolute PIXE spectra and obtain the Ge/Ga atomic concentration ratio. PIXE measurements were performed on samples A2G2, A3G3, A4G3, A5G3, A6G3, and A7G3. The minimum Ge concentration that could be detected by this technique was around $10^{21}\ cm^{-3}$. To validate the Ge concentration, secondary ion mass spectrometry (SIMS) measurements were performed on A0G3 and other reference GaN samples.

Energy Dispersive X-ray Spectroscopy (EDS) data have been collected on a Zeiss Ultra 55 scanning electron microscope (SEM) operated at 20 kV and equipped with a Flat Quad 5060F annular detector from Bruker for EDS measurements [28]. EDS measurements were performed on samples A0G3, A1G3, A3G3, A4G3, A5G3, and A6G3. The minimum Ge concentration that can be detected in our system is in the range of $5\times10^{18}$ to $10^{19}\ cm^{-3}$.

The Ge atom concentration in sample A0G3 was studied by atom probe tomography (APT) in a CAMECA Flextap system, operated in ultraviolet-laser pulsing mode at a temperature of 40 K. APT is based on the sequential field effect evaporation of individual atoms located at the surface of a needle-shaped specimen extracted and milled by focused ion beam patterning. The three-dimensional distribution of atoms and their chemical nature are obtained from a two-dimensional position sensitive detector combined with time-of-flight mass spectrometry.

Hall effect measurements were performed in the Van der Paw configuration, with the sample mounted on the cold finger of a closed-cycle He cryostat. The magnetic field was varied in the −1 to +1 T range, measured with probe located at the back side of the sample holder. Results represent the average of typically four-six measurements performed at various values of current, chosen to obtain a Hall voltage in the range of 1-30 mV. Note that the samples are designed so that there is no effect of the substrate on the Hall effect measurements. The AlN-on-sapphire templates are insulating, and the polarization difference between AlN and AlGaN generates a negative charge sheet at the substrate/AlGaN interface, which implies a depletion region propagating into the AlGaN layer. However, the AlGaN layers are thick enough to neglect the effect of the depleted area (< 5% of the total thickness) on transport measurements.

Low-temperature ($T$ = 5 K) photoluminescence (PL) spectra were obtained by excitation with a frequency-doubled solid-state laser (λ = 488 nm), with an optical power of ≈ 5 μW focused on a spot with a diameter of ≈ 100 μm. The emission from the samples was collected by a Jobin Yvon HR460 monochromator equipped with an ultraviolet-enhanced charge-coupled device camera.

## 3. Results and discussion

The Al mole fraction of the Ge-doped AlGaN samples was studied by RBS, with the results listed in table 1. As an example, the RBS spectrum of sample A7G3 is displayed in Fig. 1 together with the NDF fit that was used to calculate the Al content. The error in the estimation of the chemical composition is ±0.01 for all the samples.

The structures were further analyzed by XRD using symmetric ω–2θ scans of the (0002) reflection for all the samples and the asymmetric (−1015) reflection for selected samples. From the angular location of the symmetric reflection, it is possible to extract the lattice parameter *c*, and the strain along *c*, $\varepsilon_{zz}$. From this value, and assuming that the strain is biaxial, we calculated the in-plane strain as $\varepsilon_{xx} = -(c_{33}/2c_{13})\varepsilon_{zz}$, with $c_{ij}$ being the elastic constants of the AlGaN layers [29]. The results are summarized in Table 1 and Fig. 2(a). Let us remind here that the in-plane lattice mismatch between GaN and the AlN template is ≈ 2.6%. The AlGaN samples feel an important compressive stress at the beginning of the growth, and the mismatch relaxes progressively, as observed for binary compounds grown under metal excess [30]. The strain that remains after growth is around $\varepsilon_{xx}$ = 0.12±0.04% (see ref. [1] for a statistical analysis) and increases with the Al mole fraction of the layers, as illustrated in Fig. 2(a). This progressive relaxation process of the AlGaN layer manifests in the elongation of the (−1015) reflection towards more negative $Q_x$ and smaller $Q_z$ ($Q_x$, $Q_z$ are the reciprocal space vectors) in Fig. 2(b), which contrasts with the $Q_x$ symmetry of the reflection from the AlN substrate. These features (magnitude of the residual strain and asymmetry of the XRD reflections) do not depend on the Ge content of the layers, being present even for the more heavily doped samples [Ge] > $10^{21}$ cm$^{-3}$.

The Ge distribution in the samples was analyzed by PIXE, as illustrated by the 530×530 µm$^2$ elemental maps in Fig. 3 for samples A6G3 and A7G3. The average Ge concentration in these samples extracted from the PIXE data is in the range of 2.2-4.5×$10^{20}$ cm$^{-3}$. The images show that high Al mole fraction (x = 0.64-0.66 in the figure) leads to inhomogeneous Ge incorporation with tens-of-µm-sized areas with large clustering. In the Ge-rich regions, the Ge concentration reaches locally values around $3\times10^{21}$ cm$^{-3}$. If we keep in mind that the Ge cell temperature during the growth of these layers corresponds to a Ge incorporation in the range of 1.2-1.3$\times10^{21}$ cm$^{-3}$ in $Al_xGa_{1-x}N$ with x<0.2, these results point to a limitation in the Ge incorporation in AlGaN layers with high Al mole fraction.

To gain further insight on the nature of such Ge-rich areas, EDS/SEM studies of the samples with the highest Ge concentration have been performed. Figure 4 illustrates the results obtained on sample A6G3. An EDS map of Ge identifies Ge-rich spots [red spots in Fig. 4(a)] with a density similar to that of the Ge-rich areas identified by PIXE. In the corresponding SEM image of the layer surface [Fig. 4(b)], the Ge-rich spots appear as bright points. Increasing the magnification [Fig. 4(d)], we observe that the bright points are crystallites embedded in the semiconductor layer. Chemical analysis in Fig. 4(c) shows that the crystallites are constituted of Ge, without any trace of Al (below the noise level) and there is no perturbation of the homogeneity of the Al distribution around the crystallite. Similar results were obtained when analyzing samples A4G3 and A3G3 (x = 0.47 and 0.36, respectively), whereas no crystallites were observed in samples A1G3 and A0G3 (x = 0.12 and 0, respectively), which presented a rather homogeneous Ge distribution.

The absence of Ge inclusions in GaN:Ge was further verified by APT measurements, which give access to information at the nanometer scale. The detection of $Ge^{+}$ and $Ge^{2+}$ ions with their isotopes was homogeneous in the volume of the APT specimen, as illustrated in Fig. 5, in comparison with the distribution of $Ga^{+}$ and $Ga^{2+}$. The Ge concentration was estimated at $(9.4\pm0.5)\times10^{20}$ cm$^{-3}$. In view of these results, we conclude that Ge segregation and clustering forming crystallites is induced by increasing the Al concentration in the AlGaN alloy, probably due to the smaller Al-N bond length with respect to Ga-N and Ge-N.

The evolution of the free carrier concentration, n, and resistivity, ρ, as a function of the Al mole fraction was studied by the Hall Effect technique. The results at room temperature are presented in Table 1. For $Al_xGa_{1-x}N$ samples with $x < 0.4$, the data is plotted in Fig. 6(a) as a function of the temperature of the Ge cell. Results are compared with the case of GaN (dashed line taken from our previous work [1] and sample A0G3), which is consistent with $n \propto \exp(-E_{Ge}/k_B T_{Ge})$, where $E_{Ge} = 3.58$ eV is the thermal evaporation energy of Ge, and $k_B$ is the Boltzmann constant. For $x = 0.12$, results are approximately the same as for GaN. On the contrary, increasing the Al mole fraction to $x = 0.24$ leads to a decrease of the carrier concentration to 40% of the value for GaN, but maintains the slope of the trend, which points to the dopant level getting deeper into the band gap. For $x = 0.36$, the carrier concentration decreases further, and the variation with $T_{Ge}$ starts to deviate from the trend for GaN. This is understood as a change in the nature of the dopant, which is not only getting deeper in the bandgap but also self-compensating. This Al concentration is well below the predicted value for the onset of the DX configuration, but it could be related to the initiation of the clustering process that was described above.

For the highest Ge cell temperature used in this study, $T_{Ge}$ =1011°C (leading to $n = 1.3\times10^{21}$ cm$^{-3}$ in GaN, sample A0G3), the variation of n and ρ with the Al content is plotted in Fig. 6(b). A significant decrease in n is observed for $x > 0.24$. At the highest Al composition of $x = 0.66$, we report a resistivity of 3.8 Ωcm. For this sample, reliable measurements of n were not possible due to the high resistance.

Temperature dependent (77 K to 300 K) Hall effect measurements were carried out to determine the effective donor activation energy, $E_a$. The $E_a$ value extracted from samples grown with $T_{Ge}$ = 840 and 1011°C is plotted against the Al content in the inset of Fig. 6(b). For both $T_{Ge}$, $E_a$, increases with the Al content, reaching ≈ 40 meV for x = 0.64 and $T_{Ge}$ = 1011°C. However, this value should be taken as an underestimation of the donor energy level of isolated Ge substitutional donors. For samples with the same Al content, $E_a$ decreases for increasing doping levels, reaching almost zero as the carrier concentration approaches the Mott transition ($n \approx 1\times10^{19}$ cm$^{-3}$ for GaN [1]), which readily places carriers in the conduction band. A measurement of the activation energy of isolated Ge donors would require measurements in samples with much lower doping levels, well below $10^{17}$ cm$^{-3}$, which are unfortunately too resistive.

The normalized low-temperature ($T$ = 5 K) PL spectra of the $Al_xGa_{1-x}N$ samples with the highest doping level ($T_{Ge}$ = 1011°C) is displayed in Fig. 7(a). The spectra are dominated by the band-to-band emission, which blue shifts with increasing Al content. The PL peak position of all the samples are provided in table I, from where we can infer that, for a constant Al content, we observe a blue shift with increasing Ge incorporation due to band filling. However, as the Al content is increased, the blue shift with Ge incorporation becomes less significant due to the lower density of free carriers.

The evolution of PL Intensity $I_{PL}$ with temperature is represented in Fig. 7(b) for two of the samples (A2G3 and A6G3). For the samples with higher Ge concentration ($T_{Ge}$ = 1011°C, [Ge] ≈ $10^{21}$ cm$^{-3}$), this behavior can be well described by a one-center model, using the equation [31,32]

$$I_{PL}(T) = \frac{I_0}{1+Aexp\left(-\frac{E_{aPL}}{kT}\right)} \quad (1)$$

where $I_0$ is the PL intensity at 0 K, $A$ is a fitting parameter, and $E_{aPL}$ is the activation energy of the non-radiative recombination process. The values of $E_{aPL}$ extracted from fitting the experimental data to equation (1) are summarized as solid symbols in the inset of Fig. 7(b). For all samples, $E_{aPL}$ remains in the range of 50-90 meV, with a slight tendency to increase for higher Al mole fraction. For the samples with lower Ge concentration ($T_{Ge}$ = 840°C, [Ge] ≈ $10^{19}$ cm$^{-3}$), the thermal quenching is more complex, requiring two activation energies to get a reasonable fit to the experimental results:

$$I_{PL}(T) = \frac{I_0}{\left[1+A_1 exp\left(-\frac{E_{aPL1}}{kT}\right)\right]\cdot\left[1+A_2 exp\left(-\frac{E_{aPL2}}{kT}\right)\right]} \quad (2)$$

The resulting activation energies are presented as hollow symbols in the inset of Fig. 7(b), and they increase clearly with the Al content. The lower activation energy is probably related to exciton localization at defects or alloy inhomogeneities, which are screened in the samples with higher carrier concentration.

Figure 7(c) describes the variation of band-to-band emission energies (PL peak position $E_{PL}$) with temperature for samples with x = 0.24 and x = 0.36. For the lower doping levels, the evolution presents an S-shape behavior, with a blue shift between 70 K and 150 K. This is explained by the fact that the low-temperature emission is dominated by transitions of carriers localized in alloy fluctuations [33,34]. As temperature is increased beyond 150 K, carriers get thermally delocalized, and $E_{PL}$ approaches Varshni's law [35]. For the samples with the lowest doping level, the shift in PL peak energy at 5 K with respect to that described by Varshni's law was used to estimate a localization energy $E_{loc}$ = 23±3 meV in both cases. Such a localization energy can be justified by alloy fluctuations in the order of ±1%. When increasing the doping level, the S-shape evolution gets attenuated due to the screening of the potential fluctuations by free carriers. For carrier concentrations in the $10^{20}$ cm$^{-3}$ range, the potential fluctuations due to the alloy inhomogeneities are fully screened even at low temperatures.

## 4. Conclusion

In conclusion, we synthesized Ge-doped $Al_xGa_{1-x}N$ with Al mole fraction up to x = 0.66. We demonstrated that Ge does not induce any structural degradation in AlGaN samples with $x < 0.15$. However, for higher Al composition, we observe Ge segregation and clustering forming crystallites, probably induced by the smaller Al-N bond length with respect to Ga-N and Ge-N. Keeping the Ge concentration constant to [Ge] = $1\times10^{21}$ cm$^{-3}$, Hall effect measurements at room temperature show a gradual decrease of the carrier concentration when increasing the Al mole fraction. Layers with x = 0.64-0.66 remain conductive ($\sigma$ = 0.8-0.3 $\Omega^{-1}$cm$^{-1}$), but the carrier concentration drops to $10^{18}$ cm$^{-3}$, which implies a donor activation rate of 0.1%. From the optical point of view, the low temperature PL is dominated by near-band-edge emission. From the evolution of the PL peak energy with temperature, we observe the screening by free carriers of the localization induced by fluctuations of the alloy composition.

**Acknowledgements.** The authors acknowledge technical support by Y. Genuist and Y. Curé. This work is supported by the French National Research Agency via the GaNEX program (ANR-11-LABX-0014) and the UVLASE project (ANR-18-CE24-0014), and by the Auvergne-Rhône-Alpes Region via the PEAPLE project.

**Table I.** Description of samples under study: Al content obtained from RBS, Ge cell temperature ($T_{Ge}$), in-plane strain ($\varepsilon_{xx}$), room-temperature carrier concentration (n) and resistivity (ρ) from Hall effect measurements, PL peak location at 5 K.

| **Sample** | **Sample** | **Al mole fraction** | **$T_{Ge}$ (°C)** | **$\varepsilon_{xx}$ (%)** | **n ($cm^{-3}$)** | **ρ (Ωcm)** | **PL peak (nm)** |
|---|---|---|---|---|---|---|---|
| E3763 | A0G3 | 0 | 1011 | −0.22 | $1.3\times10^{21}$ | $2.1\times10^{-4}$ | 341.5 |
| E3743 | A1G1 | 0.12±0.01(*) | 840 | −0.10 | $1.0\times10^{19}$ | $9.8\times10^{-3}$ | 321.5 |
| E3745 | A1G2 | 0.12±0.01(*) | 928 | −0.11 | $2.1\times10^{20}$ | $1.1\times10^{-3}$ | 332.0 |
| E3746 | A1G3 | 0 .12±0.01(*) | 1011 | −0.18 | $1.2\times10^{21}$ | $2.9\times10^{-4}$ | 332.0 |
| E3747 | A2G1 | 0.24±0.01(*) | 840 | −0.11 | $6.2\times10^{18}$ | $4.7\times10^{-2}$ | 319.5 |
| E3751 | A2G2 | 0.24±0.01 | 928 | −0.14 | $6.7\times10^{19}$ | $6.4\times10^{-3}$ | 319.8 |
| E3753 | A2G3 | 0.24±0.01(*) | 1011 | −0.17 | $4.5\times10^{20}$ | $8.4\times10^{-4}$ | 316.1 |
| E3752 | A3G1 | 0.36±0.01(*) | 840 | −0.28 | $2.8\times10^{18}$ | $8.6\times10^{-2}$ | 301.0 |
| E3750 | A3G2 | 0.36±0.01(*) | 928 | −0.22 | $9.8\times10^{18}$ | $3.7\times10^{-2}$ | 302.0 |
| E3754 | A3G3 | 0.36±0.01 | 1011 | −0.18 | $5.6\times10^{19}$ | $5.8\times10^{-3}$ | 302.0 |
| E3764 | A4G3 | 0.47±0.01 | 1011 | −0.31 | $2.4\times10^{20}$ | $6.2\times10^{-3}$ | 283.0 |
| E3765 | A5G3 | 0.53±0.01 | 1011 | −0.15 | $4.9\times10^{19}$ | $3.4\times10^{-2}$ | 280.0 |
| E3782 | A6G3 | 0.64±0.01 | 1011 | −0.33 | $1.2\times10^{18}$ | 1.4 | 275.0 |
| E3783 | A7G3 | 0.66±0.01 | 1011 | −0.35 | -- | 3.8 | 272.0 |

(*) Estimation from RBS measurements in other samples.

## FIGURE CAPTIONS

**Figure 1.** RBS spectrum of sample A7G3 (Ge-doped $Al_{0.66}Ga_{0.34}N$) with the fit generated by NDF code. The signal for Ga, Al, and N at the surface are marked by arrows.

**Figure 2.** (a) In-plane strain in the AlGaN:Ge layers extracted from $\omega$–$2\theta$ scans around the (0002) reflection. (b) Reciprocal space map of sample A3G3 showing the (–1015) reflections of the Ge-doped AlGaN layer and the AlN template.

**Figure 3.** Elemental (Al, Ga and Ge) PIXE maps from samples A6G3 and A7G3. All the images represent a 530×530 $\mu m^2$ area. Al and Ga appear homogeneously distributed in the samples, whatever the Ga/Al ratio.

**Figure 4.** (a) EDS map of Ge in sample A6G3, with the corresponding SEM image (b). Note the correlation between red spots indicating high Ge concentration in (a) and bright spots on the layer surface in (b). (c) Magnified EDS map of Ge (red) and Al (blue) with the corresponding SEM image (d). The white spots in (b) are identified Ge crystals in the magnified images (c,d).

**Figure 5.** Elemental (Ga and Ge) APT reconstructions of a tip-shaped specimen from the Ge-doped GaN sample.

**Figure 6.** (a) Variation of the carrier concentration at room temperature measured by Hall effect as a function of the temperature of the Ge cell. Experimental values for Ge-doped $Al_xGa_{1-x}N$ (x = 0, 0.12, 0.24, and 0.36) are presented. The dashed line indicates the variation in the case of GaN (taken from ref. [1]). (b) Evolution of the carrier concentration and resistivity with the Al concentration in $Al_xGa_{1-x}N$ layers with [Ge] = $1\times10^{21}$ $cm^{-3}$.

Inset: Activation energy as a function of x for [Ge] = $10^{19}$ $cm^{-3}$ (hollow symbols) and $10^{21}$ $cm^{-3}$ (full symbols).

**Figure 7.** (a) Normalized low temperature ($T$ = 5 K) PL spectra of the highest doped samples ($T_{Ge}$ = 1011°C). (b) Variation of the PL intensity with temperature in samples A2G3 and A6G3. Solid lines are fits to equation (1). In the inset, variation of the activation energies of non-radiative processes as a function of the Al mole fraction, for samples with [Ge] = $1\times10^{21}$ $cm^{-3}$ (solid symbols) and [Ge] = $1\times10^{19}$ $cm^{-3}$ (hollow symbols). Note that in the case of the lower doping concentration, two activation energies are required to get a good fit to the experimental results. (c) Variation of the PL peak energy as a function of temperature in Ge-doped $Al_xGa_{1-x}N$ samples with x = 0.24 and x = 0.36 and various carrier concentration densities. Solid lines follow Varshni's equation, $E_{PL}(T) = E_{PL}(T = 0) - \alpha T^2/(T + \beta)$, with $\alpha$ = 0.590 meV/K and $\beta$ = 600 K [35].

# Figure 1

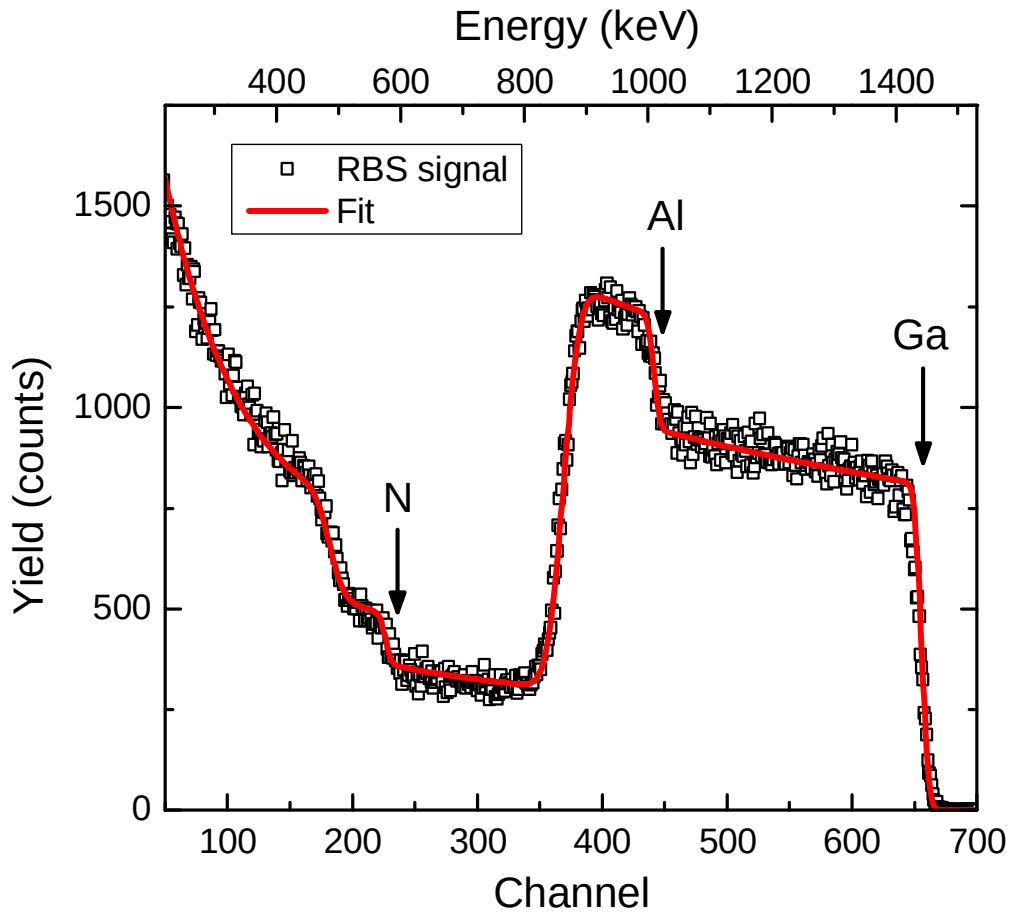

# Figure 2

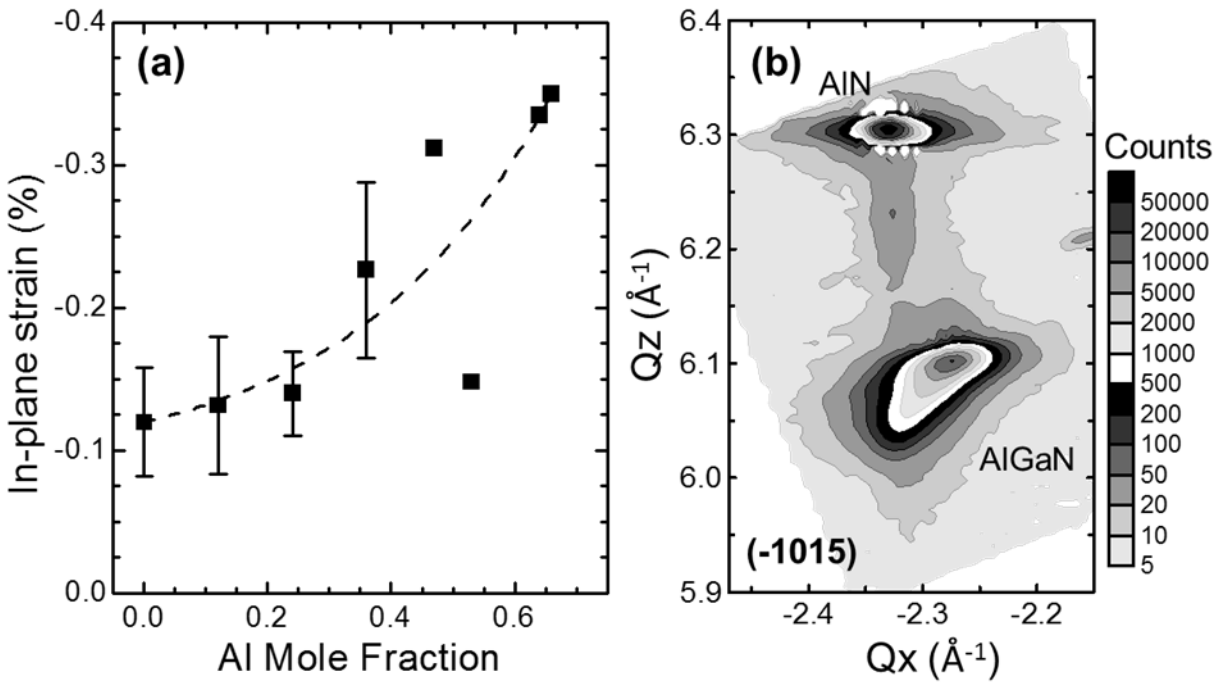

## Figure 3

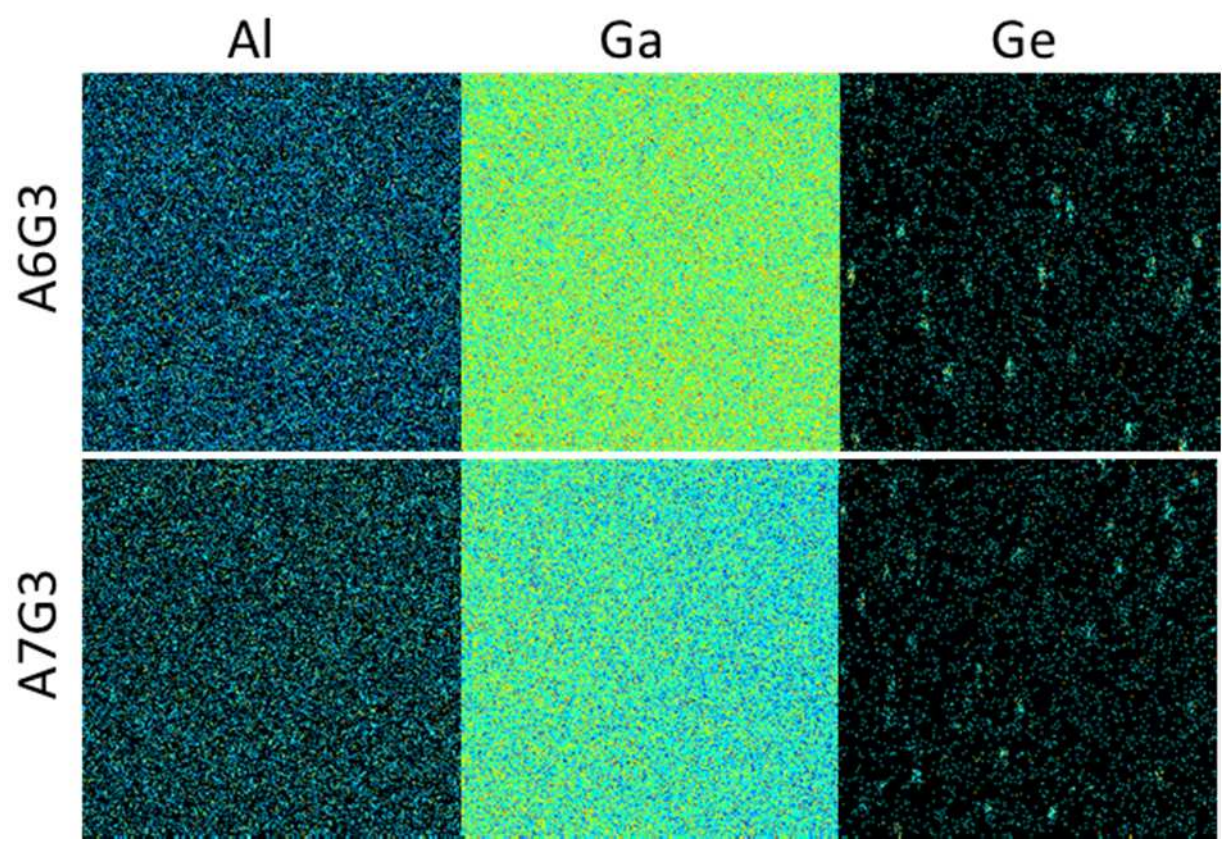

## Figure 4

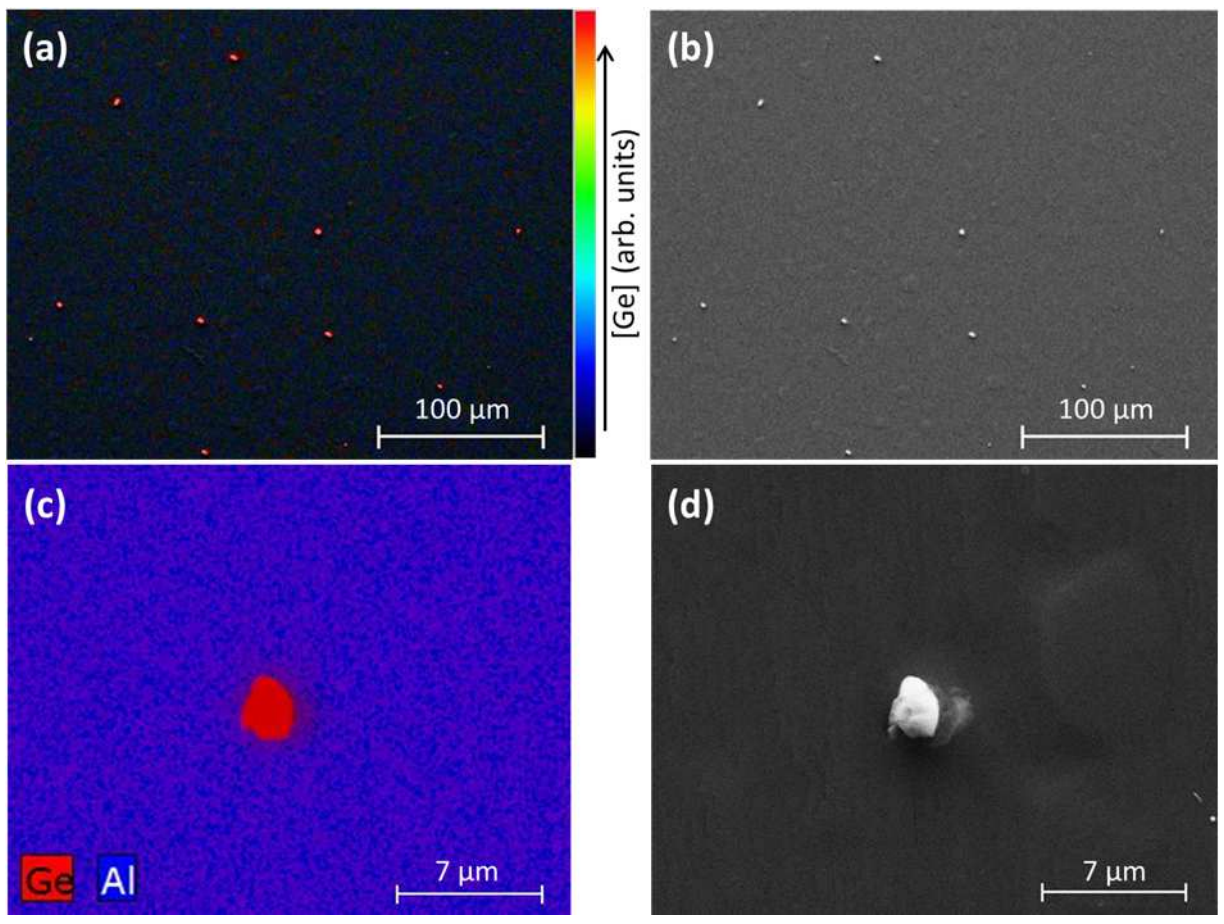

## Figure 5

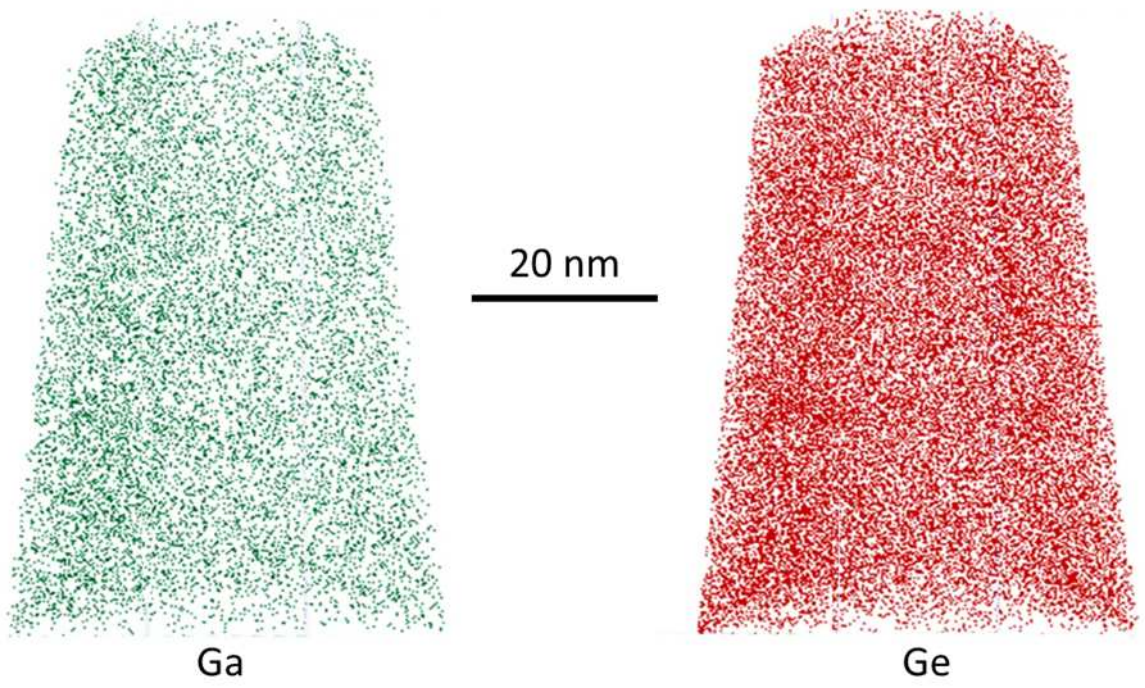

**Figure 6**

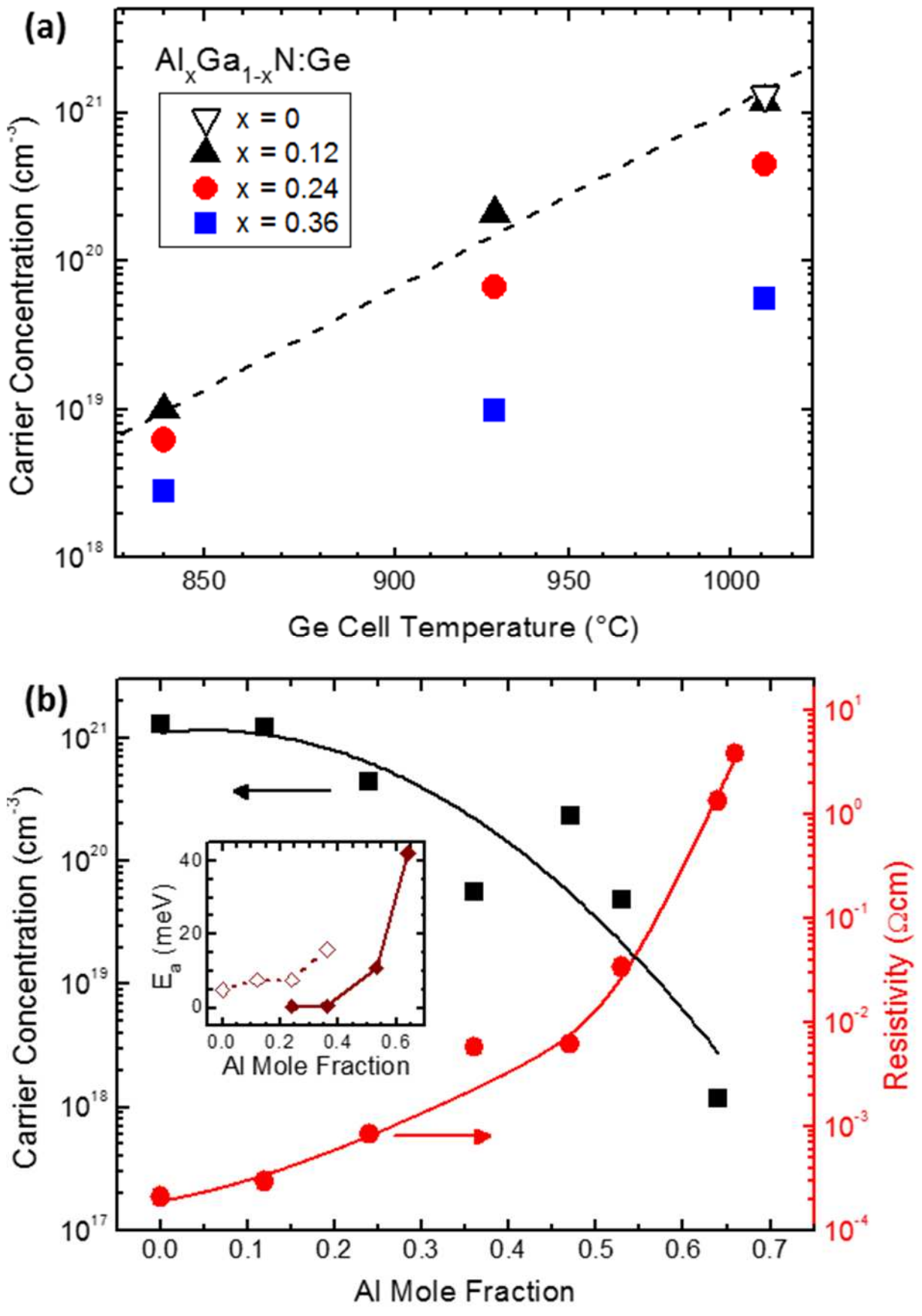

# Figure 7

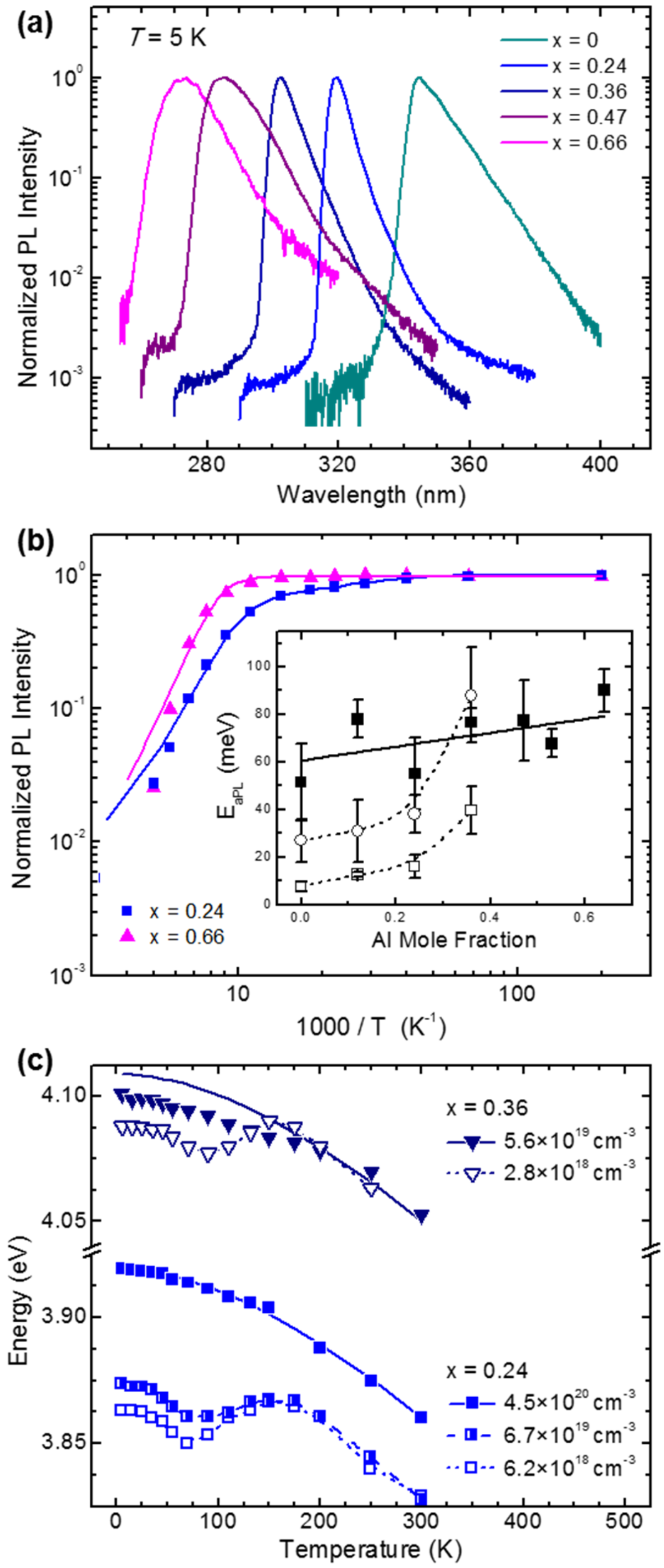